\documentclass[twocolumn]{article}

\usepackage{PRIMEarxiv}
\usepackage{cuted}
\usepackage[utf8]{inputenc} 
\usepackage[T1]{fontenc}    
\usepackage{xcolor}
\usepackage{hyperref}       
\usepackage{url}            
\usepackage{booktabs}       
\usepackage{amsfonts}       
\usepackage{amsmath}       
\usepackage{nicefrac}       
\usepackage{microtype}      
\usepackage{lipsum}
\usepackage{fancyhdr}       
\usepackage{graphicx}       
\usepackage{subcaption}       
\usepackage{wrapfig}  
\graphicspath{{media/}}     

\title{Three Dimensional Effects on Proton Acceleration with Grooved Hydrocarbon Targets
}

\author{
  Imran Khan$^\dagger$, Mohammad Yasir, Vikrant Saxena*\\
  Department of Physics \\
  Indian Institute of Technology Delhi, \\
  New Delhi, India\\
  \phantom{phantom}\\
  $^\dagger$ \textit{Current Affiliation:}
  Anubal Fusion Private Limited,\\
  Hyderabad, India\\
 \texttt{*} vsaxena@physics.iitd.ac.in\\
}

\begin{document}
\begin{strip}
\maketitle

\begin{abstract}
    Recently, using two-dimensional particle-in-cell simulations, it has been demonstrated that in laser based proton acceleration with micro-structured targets, a single rectangular groove on the target front offers significant proton cut-off enhancement with linearly polarised laser pulses. In the present work, three-dimensional investigations are carried out to identify notable differences between cylindrical and cuboidal groove geometries both of which correspond to a rectangular groove in a two-dimensional case. In particular, a waveguide model is employed to analyse the effect of the groove geometry and extensive three-dimensional particle-in-cell simulations are performed to demonstrate the distinct behaviour of laser pulse and electrons for cylindrical and cuboidal grooves. Further, the effect of a circular polarisation of the incident laser pulse on the spectra of accelerated protons is studied. It is shown that contrary to our initial expectations, cylindrical symmetry and circular polarisation do not play well together and cause as much as $15\%$ decay in proton cut-off energies as compared to the case of cylindrical symmetry and linear polarisation.
\end{abstract}
\end{strip}


\section{Introduction}
\label{sec_intro}
    The features of coherent acceleration \cite{veksler_coherent} are easily realised during the interaction of an intense, typically femtosecond laser pulse with a plasma target, yielding electrons and ions with energies in the MeV range. Owing to reduced financial and environmental costs, smaller length scales, and larger acceleration gradients, such acceleration setups draw significant attention with applications to proton radiography and therapy \cite{Talamonti_Proton_Radio,Johnson_Radio_Review,Mackinnon_Proton_Diagnostic}, studies and production of warm dense matter \cite{Roth_WDM,Patel_WDM}, fast ignition studies \cite{Fernandez_fast_Ignition,Roth_Fast_Ignition}, and other problems in applied and research fields.

    The typical approach to the acceleration of protons through such setups involves as a first step, the displacement of a large number of electrons, which necessitate the movement of positively charged ions to maintain charge neutrality. Acceleration from both the front and rear surfaces has been studied\cite{Fuchs_RSA_FSA}. The latter has received major attention via the Target Normal Sheath Acceleration Mechanism \cite{Wilks_TNSA} due to its relative ease and experimental feasibility \cite{Scullion_Laser_param,Bagnoud_PHELIX}. This comes at the cost of drawbacks in the beam quality, primarily including the saturation of proton cut-off energy \cite{Keppler_TNSA_Limit}, large angular spread, and strongly polyenergetic features in the output beam. Over the past several years, multiple approaches have been proposed to push the frontiers of proton acceleration using factors like target micro-structuring \cite{Grandvaus_MST}, nano-structuring \cite{Vallieres_Nanowire}, and external field stimuli \cite{Khan_2024_MagField}. 
    
    Noticeable success has been achieved on this front via hybridising/cascading other acceleration mechanisms with TNSA. This has been done either via target structuring \cite{Grandvaus_MST,Khan_Groove_Compare}, or by careful control of laser-target parameters \cite{ziegler_cascaded_2024}. In the former case, structures on the target front alter laser-plasma interaction which can lead to enhanced energy deposition. For instance, micron-level grooves can provide optical confinement, leading to an increase in laser intensity inside plasma channels as studied in ref \cite{Ji_Laser_In_Channel}. This approach appears appealing owing to the relative ease that modern etching/drilling methods provide for precise target creation. In the latter case, ensuring optimum target thickness and timing the laser pulse so that RIT occurs at laser peak arrival \cite{Dover2023} allows for the generation of spectrally separated high-energy protons.
    
    On the micro-structuring front, it has been demonstrated that a single micron-sized groove on the target front improves hot electron generation via direct laser acceleration, which can push proton cut-off energies to beyond the 100 MeV regime \cite{khan_2025_RIT}. These grooves with overdense walls behave similar to plasma channels/waveguides and variations on such a structure to perform pulse guiding \cite{Chiou_optical_guiding} and particle acceleration \cite{Tajima_PFA} have been explored. In previous studies, the groove was rectangular in shape and studied in two-dimensions, which suppresses transverse geometrical effects and overestimates proton cut-off energy \cite{Humieres_2D_Overestimate,Sgattoni_2D_overestimate}. Furthermore, circularly polarized lasers have demonstrated the generation of megagauss level axial magnetic fields inside plasma channels \cite{kim2002axial}, which can alter acceleration dynamics \cite{Mishra2024}. Thus, their interaction with structured targets emerges as a natural question. 
    
    In the text that follows, the simulation setup is outlined in section \ref{sec:sim_setup}. After a brief analysis in section \ref{sec:analysis}, it is demonstrated that a cylindrical groove has relative advantages due to enhanced self-focussing and forward momentum transfer when a linearly polarised laser beam is employed (section \ref{sec:linear_polarisation}). It is shown that despite higher peak electron energies, the cuboidal groove loses the battle to the cylindrical one with a linearly polarised pulse when it comes to proton cut-offs, which can reach as high as $\approx 68$ MeV for the latter case. Simultaneously, drastic disadvantage in circular polarisation is shown, even as cylindrical symmetry offers enhanced alignment with incident gaussian beam. Contrary to expectations, it is discussed how reduced electron temperature and forward momentum transfer lead to decay in proton cut-off energy for this case. This is done in section \ref{sec:circ_pol}, followed by discussions and conclusion in section \ref{sec:conclude}. Together, these results are of significant interest to the ion acceleration community, pushing the boundaries of what is possible with a contemporary laser-plasma setup.

\section{Simulation Setup}
\label{sec:sim_setup}
    The results in this text are inferred from three-dimensional particle-in-cell simulations performed via the open-source massively parallel code EPOCH \cite{Arber_EPOCH}. Our simulation domain extends $50\ \mu m$, $15\ \mu m$, and $15\ \mu m$ in x, y, and z directions, respectively, with $2500$, $375$, and $375$ cells, giving us a resolution of 20 nm in the longitudinal and 40 nm in the transverse directions.

    Two overcritical targets are considered, corresponding to the cases of cylindrical (CG, synonymously called circular groove in this text) and cuboidal (RG, synonymously called rectangular) grooves. These targets have a density corresponding to that of fully ionised polyethylene $((C_2H_4)_n)$, which has a mass density of $\rho = 0.93$ g cm$^{-3}$, leading to number densities of  $\approx 23\ n_c,\ 46\ n_c, \text{and}\ 184\ n_c$ for Carbon, proton, and electron species, respectively. These are initialized with 4, 8, and 12 particles per cell. For each case, the target is located in the range given by $x \in (0\ \mu m, 7\ \mu m)$, $y \in (-5\ \mu m,5\ \mu m)$, and $z \in (-5\ \mu m, 5\ \mu m)$. The grooves on the target front extend $6.98\ \mu m$ in the longitudinal (laser propagation) direction with a $20$ nm rear wall. The diameter of the CG case is same as side length in the RG case, carrying height and width of $3\ \mu m$.
    
    Incident on the plasma is a laser pulse of $0.8\ \mu m$ wavelength, with Gaussian profiles in both space and time (characteristic widths of $3\ \mu m$ and $40$ fs, respectively), and a peak intensity of $5.5 \times 10^{20}$ W cm$^{-2}$. At this intensity, radiation reaction effects \cite{Mishra2022} will not affect dynamics. These parameters are similar to experiments conducted \cite{Scullion_Laser_param} at Rutherford Appleton Lab (RAL), STFC, UK, with the GEMINI Ti:Sapphire laser.

\section{Analysis}
\label{sec:analysis}
    The dynamics of electrons inside the grooves may be analysed using the waveguide model proposed by Shen \cite{shen1991plasma}. This was performed in cylindrical coordinate system, but the underlying physics may be extrapolated to a rectangular system as well with modifications to the field distributions inside. In the following subsections, we set up the waveguide model to analyse electron dynamics and expected electron energies inside the groove before the rear wall effects start to set in. The dynamics beyond this point have been studied by Zou et al. \cite{zou2017laser}. Essentially, it is seen that the dragged out electrons from the groove wall are primarily responsible for setting up the rear accelerating field while the rear wall electrons constitute the return current. 

    In line with previous results, the electron temperatures for grooved targets lie well above those of plain targets, at least twice the ponderomotive scaling \cite{Wilks_TNSA} ($T_h = 0.511 \left(\sqrt{1+a_0^2}-1\right)$) and significantly higher than Beg's experimental fit \cite{beg1997study} or Haines' relativistic model \cite{haines2009hot}. This higher temperature is primarily responsible for enhancing the rear side sheath field set up due to TNSA at the rear wall. In our simulations, at the instant when the groove wall electrons have just started to interact with the rear wall, linear polarisation shows a temperature of $13.64$ MeV and $12.04$ MeV for RG and CG cases. For circular polarisation, these temperatures drop significantly and are $7.64$ MeV and $7.83$ MeV, respectively.
    
    \subsection{The Waveguide Model}
    \label{subsec:cyg}
        \begin{figure}
        \centering
            \includegraphics[width=\linewidth]{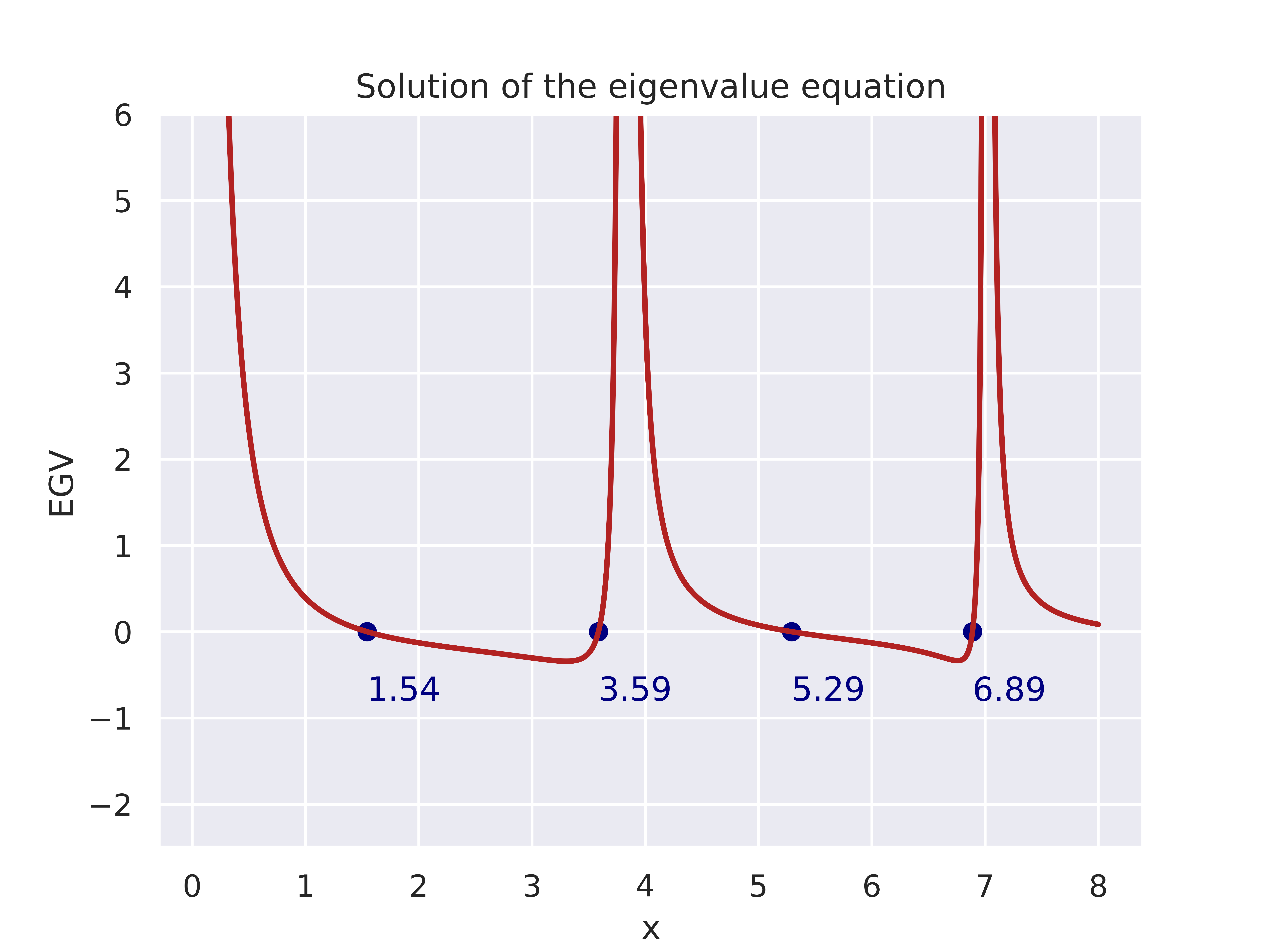}
            \caption{Graphical solution of the eigenvalue equation.}
            \label{fig:egv}
        \end{figure}
        The cylindrical waveguide has been well-studied in references \cite{shen1991plasma,Yi2016,Yi2016PoP}. The electric field components excited in a cylindrical waveguide may be calculated from the Hertz potential formalism and are given as:
        \begin{align}
            E_x &= E_{x0} J_1(Tr) \sin(\phi) e^{-jk_x x} + \text{c.c.}, \\
            E_r &= j \frac{k}{T} E_{x0} 
                \left[ \frac{J_1(Tr)}{Tr} - \frac{k_x}{k} J_1'(Tr) \right] 
                \sin(\phi) e^{-jk_x x} + \text{c.c.}, \\
            E_\phi &= j \frac{k}{T} E_{x0} 
                \left[ J_1'(Tr) - \frac{k_x}{k} \frac{J_1(Tr)}{Tr} \right] 
                \cos(\phi) e^{-jk_x x} + \text{c.c.}.
        \end{align}
        
        The longitudinal field inside the waveguide is smaller than the transverse ones by a factor of $(T/k)^2 \ll 1$. Thus, most energy is coupled into transverse modes and we have average peak acceleration gradient of $a_{0m} \approx \frac{x_1\lambda_0}{2\pi r_0}$, where $x_n$ is the n$^\text{th}$ root of the eigenvalue equation obtained by the continuity of transverse fields at the groove walls \cite{shen1991plasma}. For our case, $x_1 = 1.54$ (see figure \ref{fig:egv}). Thus, maximum electron energy is roughly $\epsilon_m \approx qE_mL_c \approx 59.16$ MeV for linear polarisation, which is in good agreement with simulation result of $\approx 61.59$ MeV. 
        
        For circular polarisation, the acceleration process is expected to remain almost similar. In our simulations, we have reduced the electric field intensity for the two components of the circular polarisation (by a factor of $\sqrt{2}$) so that the overall effect is comparable to the linear polarisation case. With this factor in mind, the simulation yields $\approx 33.36$ MeV cut-off for this case, which is comparable to the $59.16/\sqrt{2} = 41.83$ MeV estimate. These estimates use the groove depth of $L_c=6.98\ \mu m$ instead of the de-phasing length $L_d$ since the latter outweighs the former around five times. Furthermore, we are comparing the cut-off energies at the time instant where the electrons have just started to interact with the rear wall to minimize the effect of the latter in this calculation.

\section{Simulation Results}
\label{sec:sim}
\begin{figure}[!ht]
        \centering
            \includegraphics[width=\linewidth]{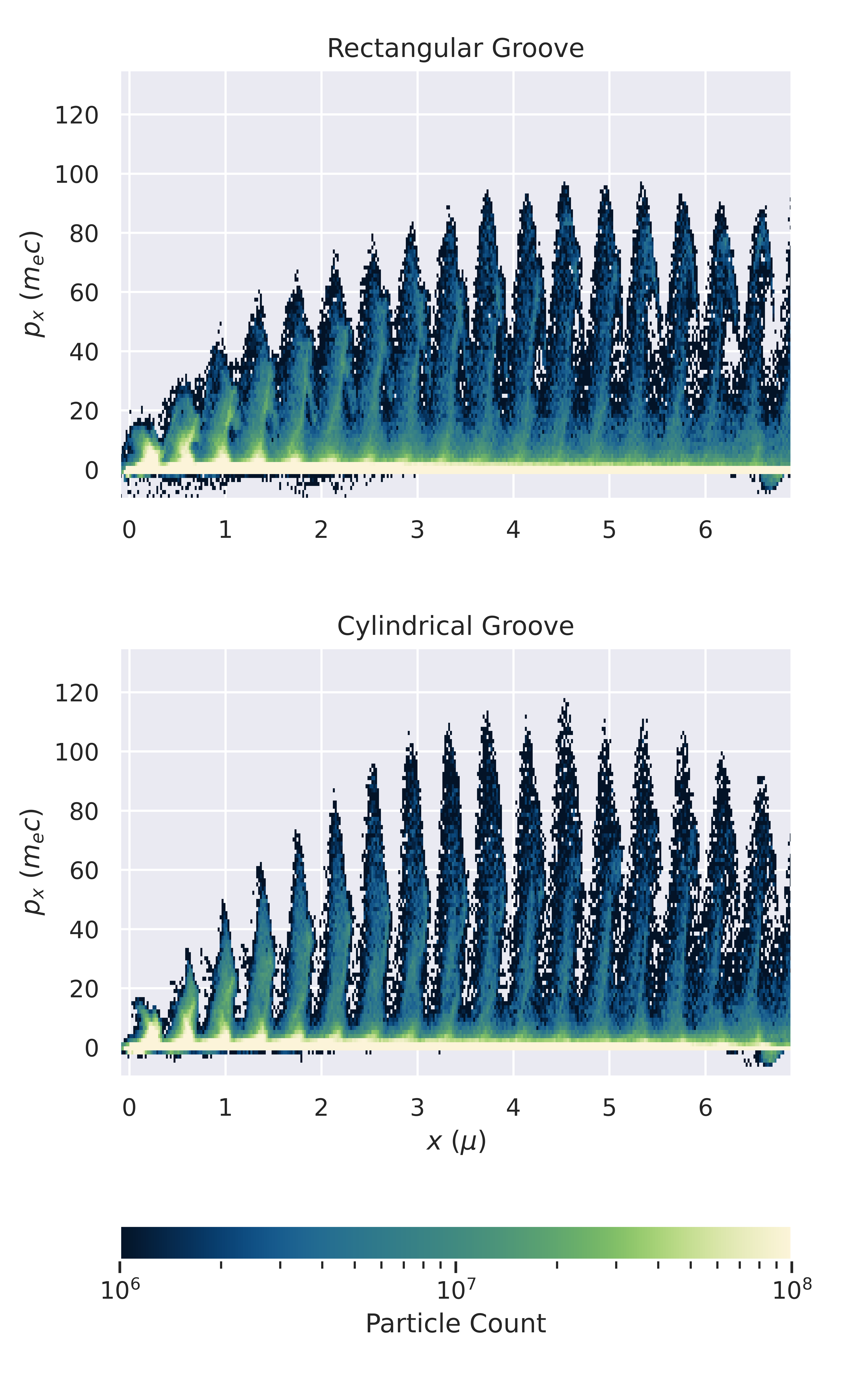}
            \caption{Electron phase space distribution at t = 60 fs for (top) rectangular groove and (bottom) cylindrical groove.}
        \label{fig:fxpx}
        \end{figure}

    The interaction of the intense laser pulse with the side walls of the groove extracts, in either geometry and polarisation, bunches of electrons with a separation of $\lambda/2$ in the phase space (plotted in figure \ref{fig:fxpx}), in line with previous reports by \cite{zhu2022bunched,ji2016towards}. This is the typical behaviour of electrons in a DLA setup and these electrons then move in phase with the laser field before reaching the target rear and inducing the TNSA sheath field. Thus, hot electron generation directly affects the performance of protons in a typical ion acceleration setup. In subsections \ref{sec:linear_polarisation} and \ref{sec:circ_pol}, we discuss our results further.

    \subsection{Linear Polarisation}
    \label{sec:linear_polarisation}
        \begin{figure}[!ht]
            \centering
            \includegraphics[width=\linewidth]{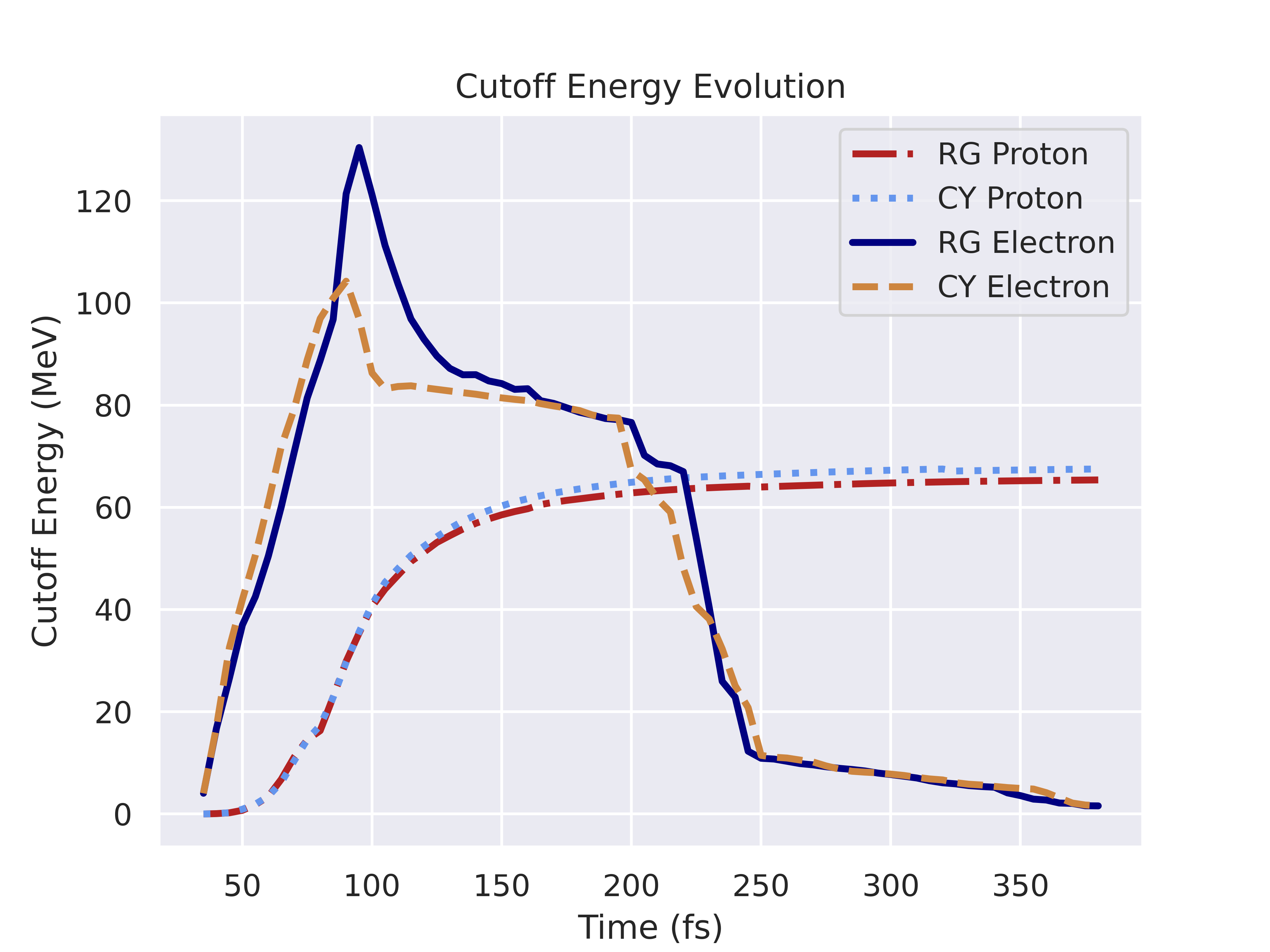}
            \caption{cut-off energy evolution for each species for linear polarisation.}
            \label{fig:cut-off_evolve}
        \end{figure}
        Figure \ref{fig:cut-off_evolve} shows the evolution of the cut-off energies for each of the species and grooves with a linearly polarised pulse. It is observed that the maximum electron energy is dominant for the case of rectangular groove, especially in the regions of $85$ fs to $175$ fs. Even so, the proton cut-off energy is sustained at slightly higher values for the cylindrical groove after the 100 fs mark. It appears surprising that electrons, which are primarily responsible for ion acceleration in such setups attain more energy with a rectangular groove, but protons do so for the CG case. The answer is attained by considering the following factors.
        
        We look at the momentum distribution inside the groove. It may be seen that the CG target imparts more longitudinal momentum to the plasma electrons as compared to the RG case. This distribution shows similar trends near the start of the target (up to $\approx 2.5\ \mu m$), but the CG case shows slightly higher peaks beyond this point. That is, even though the maximum energy imparted to electrons is higher for the RG case, the CG case has higher number of forward-directed electrons with higher energy, which leads to enhanced proton acceleration.
        
        \begin{figure}[!ht]
        \centering
            \centering
            \includegraphics[width=\linewidth]{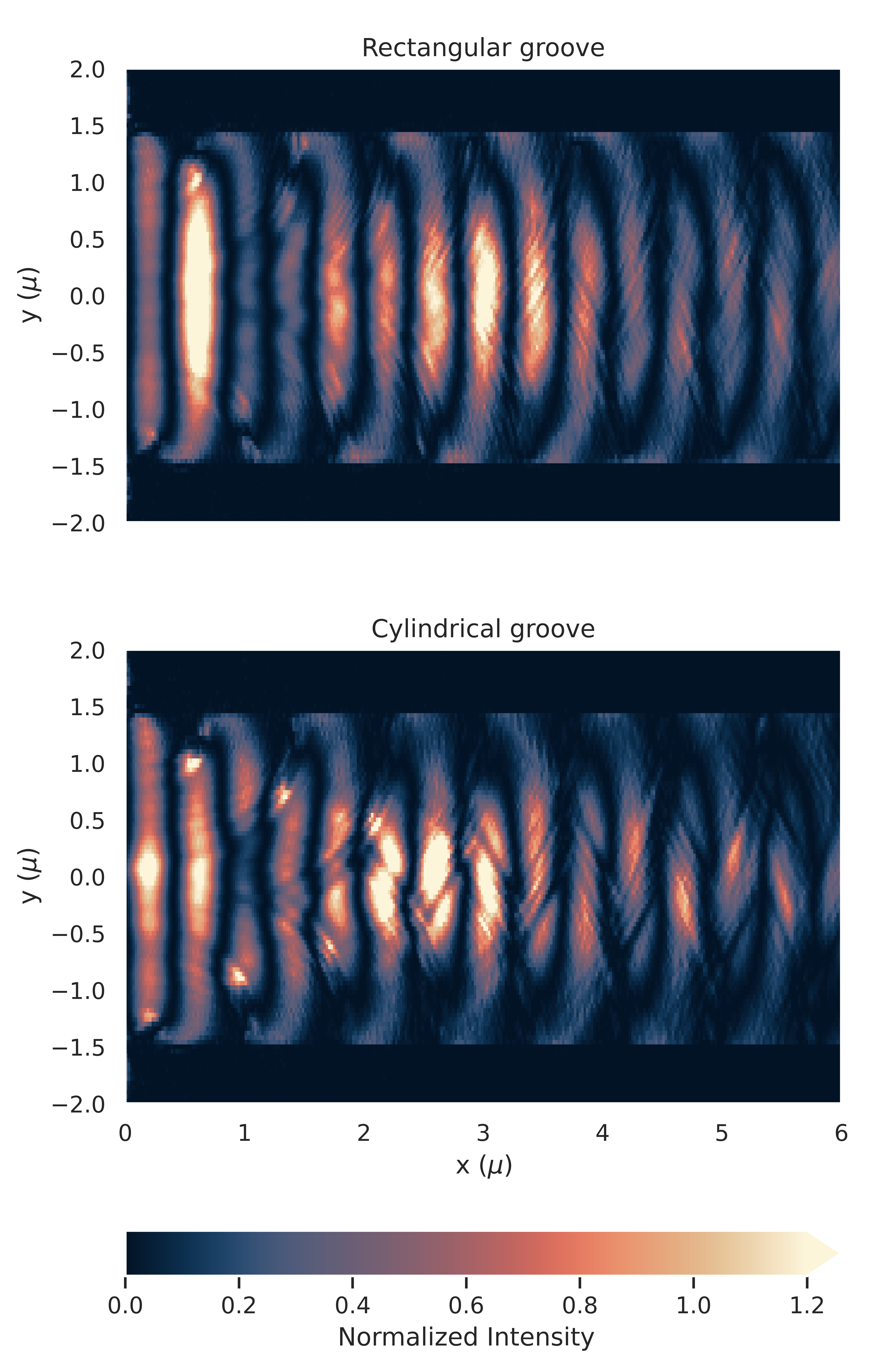}
            \caption{Laser intensity distribution normalized with respect to maximum free space intensity value at t = 60 fs for (top) rectangular and (bottom) cylindrical grooves. The colourbar at the bottom is saturated and higher values are seen in cylindrical case.}
        \label{fig:iy_12}
        \end{figure}
        This increased longitudinal momentum may be attributed to the role that the groove geometry plays. Previous studies have demonstrated that laser energy gets redistributed inside a plasma microtube \cite{ji2016towards} or groove \cite{khan_2025_RIT}. With overdense plasmas surrounding them, the grooves act like optical waveguides, leading to mode conversion, which redistributes laser pulse energy and creates localized regions of strong intensity. Assuming that $I$ and $I_0$ correspond to laser intensity with and without plasma, respectively, the intensification factor $\eta = I/I_0$ inside the groove is above 2.4 and 1.9 for CG and RG, respectively. This value is 3-4 times less compared to what has been reported in 2D simulations previously \cite{Khan_Groove_Compare}. The degree and location of intensification is significantly affected by the groove geometry and we notice, from figure \ref{fig:iy_12}, that this intensification occurs more strongly and for a comparatively larger region in the CG case. The RG case has a small region of high intensity near the very beginning of the groove but subsequent regions do not present spatially sustained intensification.
    
        \begin{figure}
        \centering
            \includegraphics[width=\linewidth]{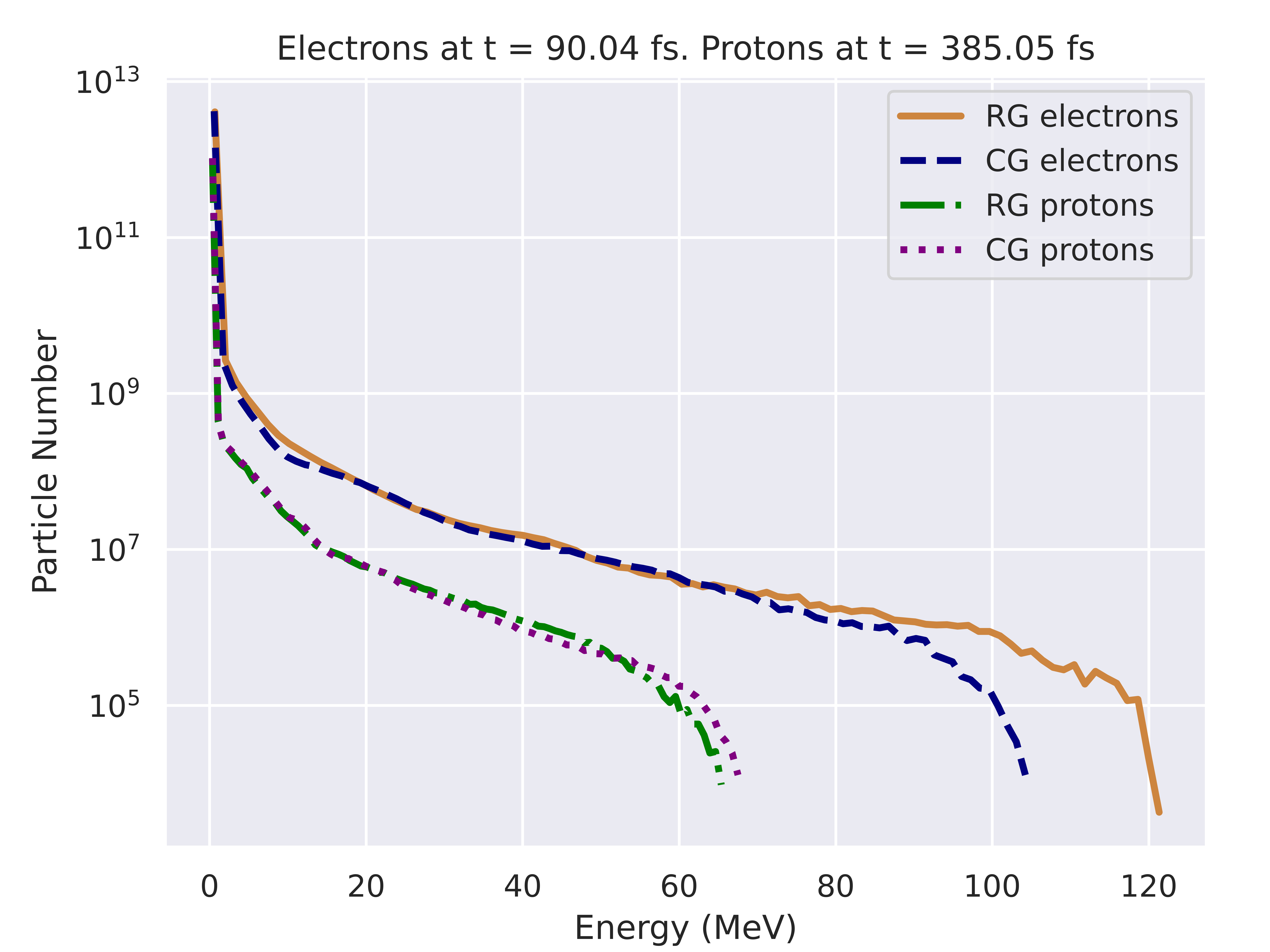}            
            \caption{Particle energy distribution for electrons at t = 90 fs and protons at t = 385 fs.}
            \label{fig:dnde}
        \end{figure}
        \begin{figure}
            \centering
            \includegraphics[width=\linewidth]{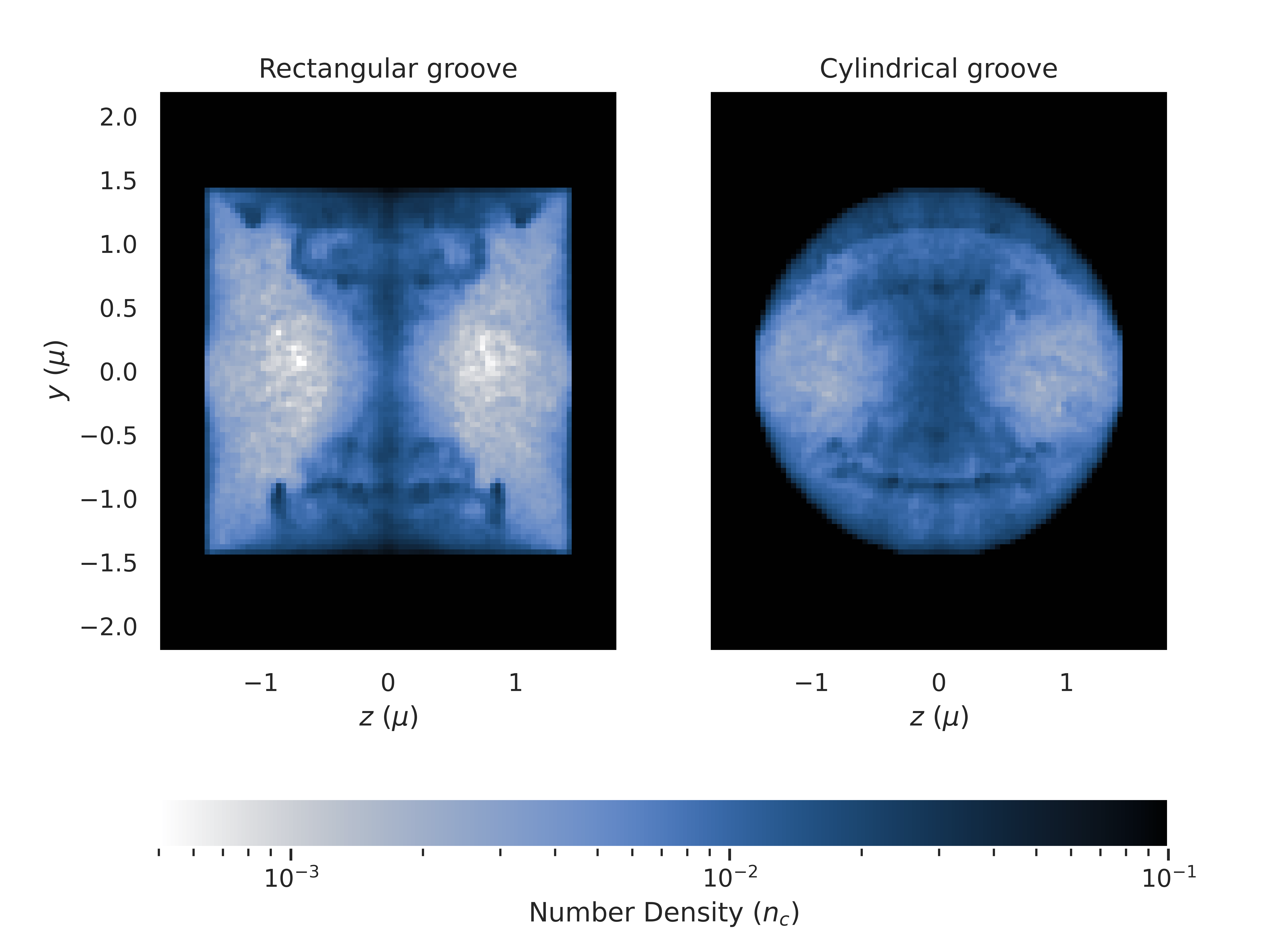}
            \caption{Longitudinal average of electron number density at t = $75$ fs for rectangular (left) groove and cylindrical groove (right) with linear polarisation.}
            \label{fig:av_nd}
        \end{figure}
        The particle energy spectrum is shown in figure \ref{fig:dnde}. Another reason why the proton cut-off is higher for the cylindrical groove may be understood by looking at the longitudinally averaged electron number density. We see that electron extraction occurs in such a way that a circular high-density front is seen moving towards the axis (figure \ref{fig:av_nd}). This leads to an on-axis maximum that is missing for the rectangular groove (electron extraction is planar); a phenomenon that has been demonstrated to be beneficial for ion acceleration \cite{Valenta2024}. This maximum starts forming around $t = 125$ fs and the proton distribution is quite similar for the two grooves until this point.

    \subsection{Circular Polarisation}
    \label{sec:circ_pol}
\begin{figure}
            \centering
            \includegraphics[width=\linewidth]{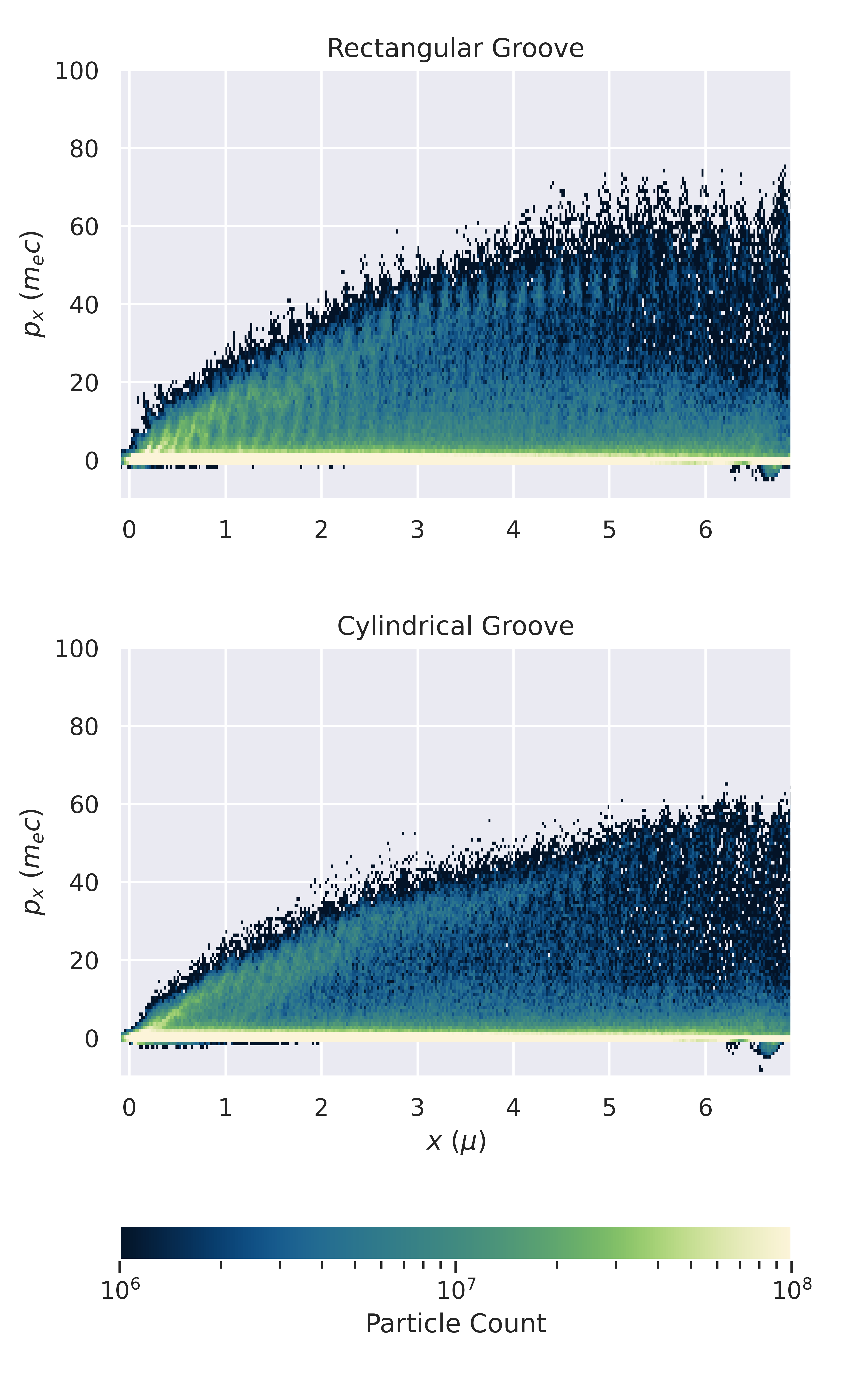}
            \caption{Electron phase space distribution at t = 60 fs for (top) rectangular groove and (bottom) cylindrical groove with circular polarisation.}
            \label{fig:d2ndxdpx_circular}
        \end{figure}

        There are certain differences in electron and proton dynamics when switching to a circularly polarised laser pulse. Electron extraction now occurs in two planes simultaneously and each direction receives only half the energy. The number density fronts are now no longer planar; instead, they take a helical shape. This type of helical electron trajectory is well established in literature \cite{sharma_cp_mva}. This helical motion also means that the momentum contribution to electrons is not dominantly longitudinal (see figure \ref{fig:d2ndxdpx_circular}). Indeed, while some structuring is visible in the $x-p_x$ phase space distribution, the overall distribution is diffuse. 

        \begin{figure}
        \centering
            \includegraphics[width=\linewidth]{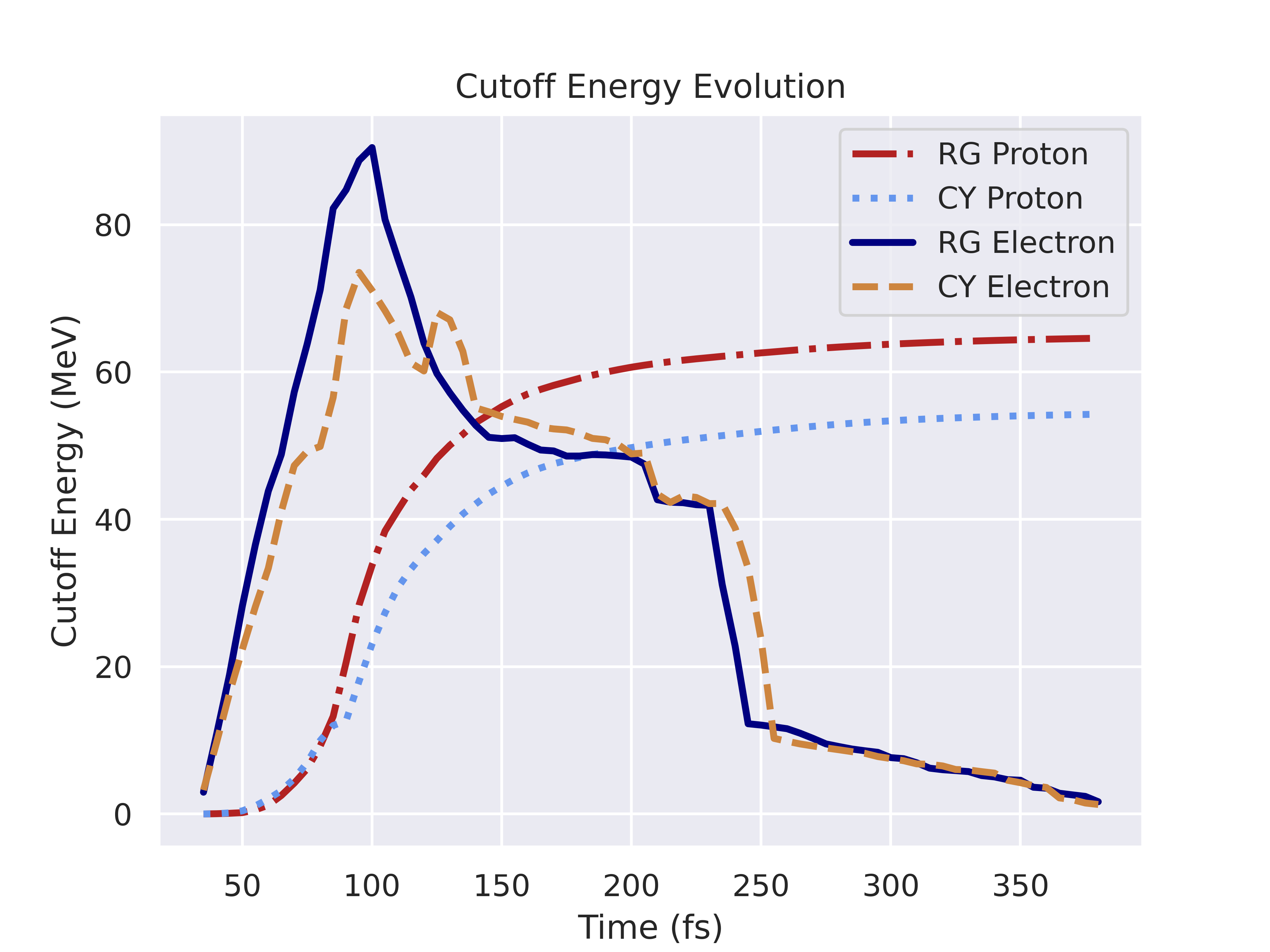}
            \caption{Evolution of cut-off energy of species for circular polarisation.}
            \label{fig:cut-off_overlay_circular}
        \end{figure}
        \begin{figure}
        \centering
            \includegraphics[width=\linewidth]{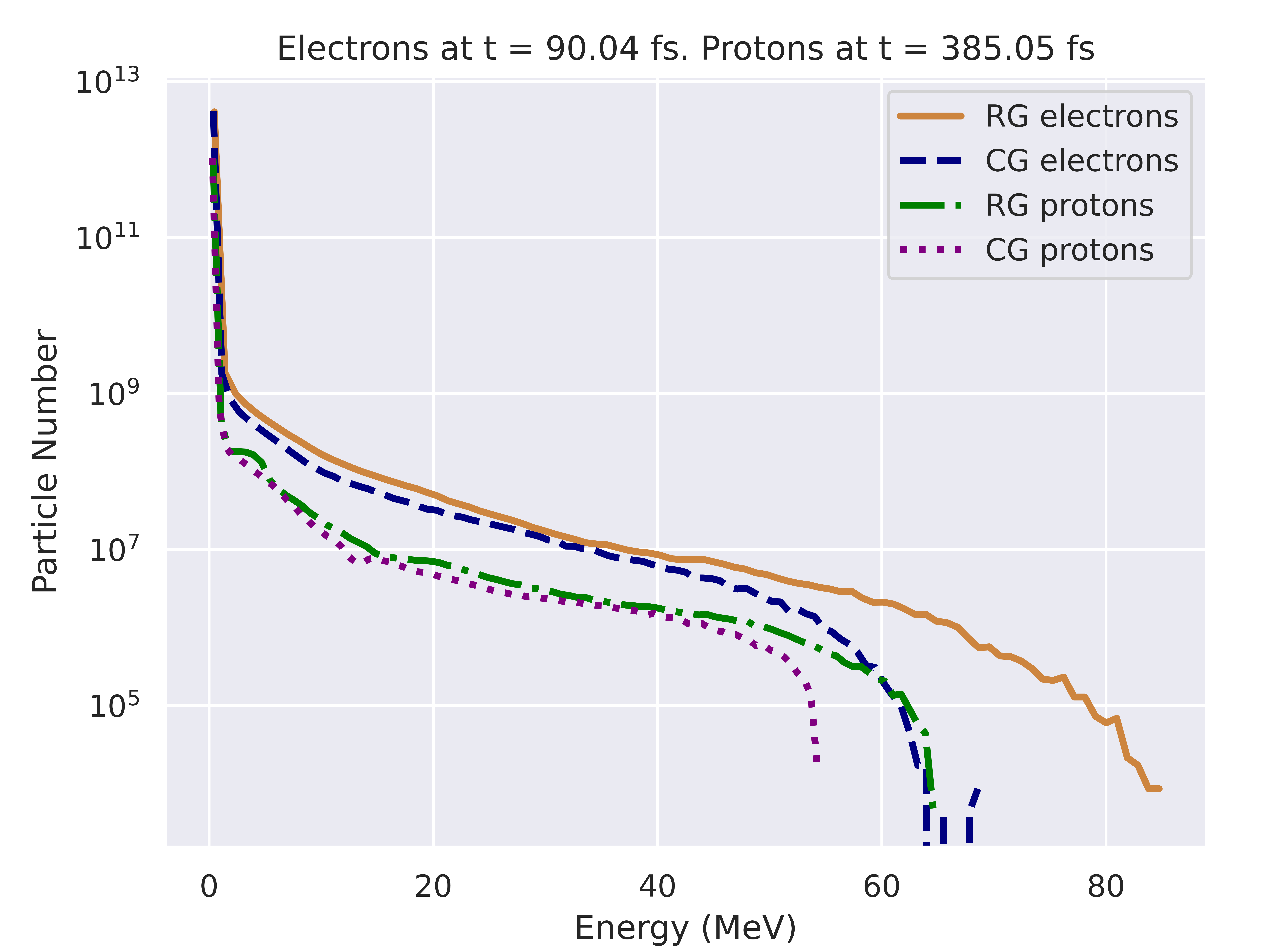}            
            \caption{Particle energy distribution for electrons at t = 90 fs and protons at t = 385 fs with circular polarisation.}
            \label{fig:dnde_circular}
        \end{figure}
\begin{figure}
            \centering
            \includegraphics[width=\linewidth]{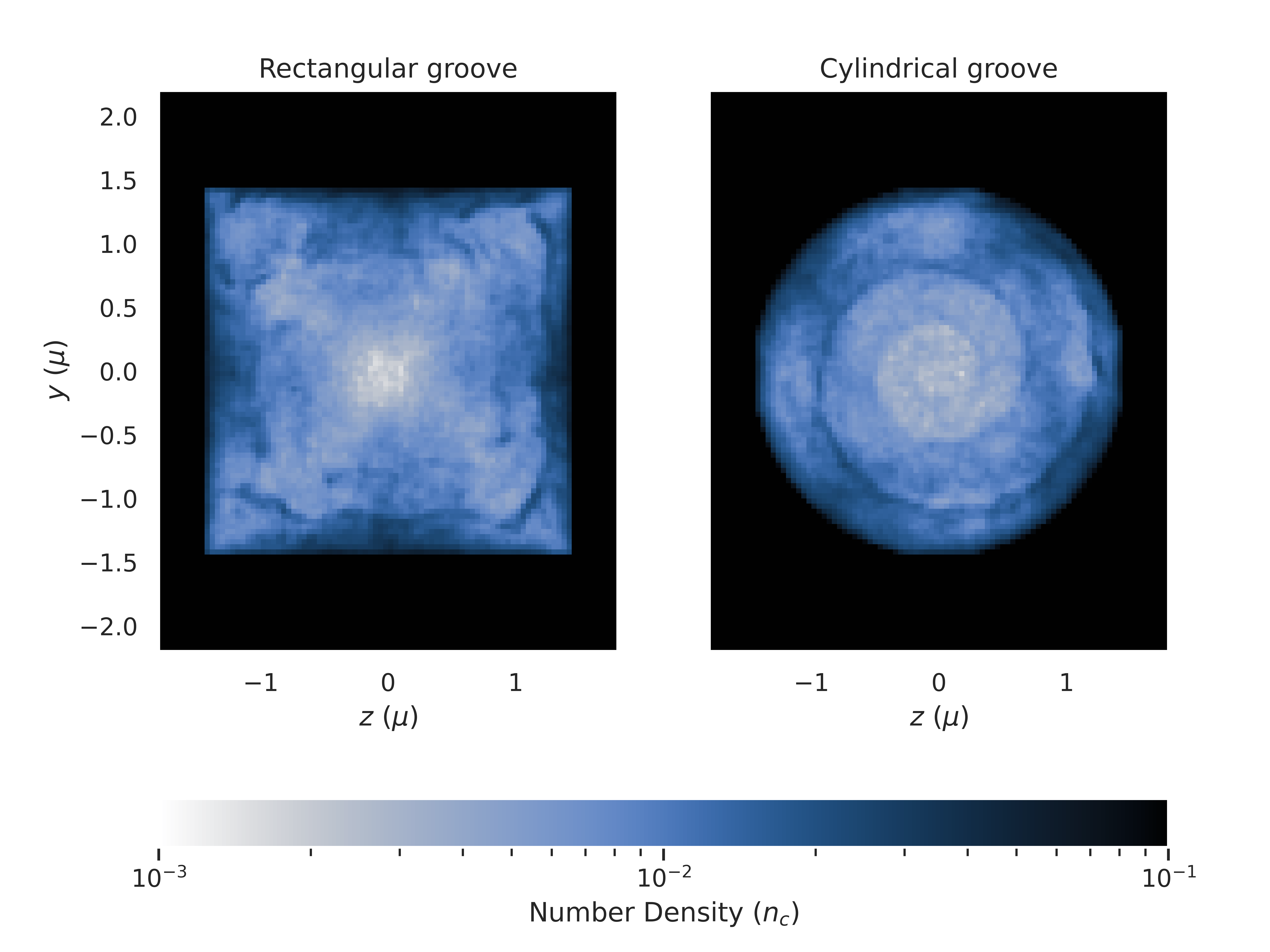}
            \caption{Longitudinal average of electron number density at t = $80$ fs for rectangular (left) groove and cylindrical groove (right) with circular polarisation.}
            \label{fig:av_nd_circular}
        \end{figure}

        As before, the maximum electron energy remains higher for RG than CG case for an extended period of time. The cut-off energy evolution is plotted in figure \ref{fig:cut-off_overlay_circular}. Sustained higher values of maximum electron energy is seen with the RG case, this time starting in the very beginning of the simulation. However, the proton cut-off trend is reversed as compared to the linear polarisation and this time, despite the inherent geometrical symmetry and alignment of the CG with a circularly polarised laser, the performance is significantly worse. As demonstrated from figure \ref{fig:dnde_circular}, there is almost a $15\%$ decay in proton cut-off energy of the CG when switching to a circular polarisation. Note that we still notice the on-axis density maximum for electrons as discussed previously (helical extraction this time, see figure \ref{fig:av_nd_circular}). Clearly, this maximum is unable to compensate for the inadequacies that arise with this polarisation.
        
        The inadequacies we speak of are primarily two: firstly, as demonstrated, the momentum distribution this time is not dominant in the longitudinal direction. Indeed, for the few electrons whose direction cosines for the forward direction fall within the $13^0$ mark, the energy is quite low for the CG case while the RG case yields larger number of electrons moving within this angular range and imparts to them, higher energy. Secondly, as has been discussed earlier, circular polarisation at normal incidence strongly suppresses hot electron generation, leading RPA to dominate over TNSA at any intensity \cite{Macchi_RPA_TNSA_Compare}. The structured target in our case relies on electron generation via DLA inside the groove. These electrons then enhance the TNSA process on the rear side wall. Since hot electron generation is less, the TNSA process is suppressed. Furthermore, the RPA process is also suppressed since its onset occurs at a minimum threshold intensity in the range of $10^{20}$ W.cm$^{-2}$ (calculated in 2D by Schmidt et al. \cite{schmidt_rpa}). In 3D, this threshold is expected to increase further. The laser pulse we have used sits exactly at this threshold intensity. Thus, proton cut-off energy is lowest this time around.
    
\section{Discussions and Conclusion}
\label{sec:conclude}
    Plasma microchannel targets aren't entirely novel and some investigation has been performed in this direction. Most of those, however, have focused on channels created inside a metallic plasma with a hydrocarbon layer at the rear side. The reader is referred to the work of Zou et al. \cite{zou2017laser} which utilises a waveguide model to analyse a microchannel target not unlike ours in terms of geometry; ours is, however, an entirely plastic based target. Some parallels have yet been drawn.

    Through previously established analytical work and novel fully three-dimensional PIC simulations, we have comprehensively explored this interesting regime and demonstrated the advantage of a cylindrical groove irradiated by a linearly polarised laser pulse. For circular polarisation, the suppression of hot electrons and the diffuse nature of $x-p_x$ distribution function is shown to provide reduced proton cut-off energies. The minor advantage offered by cylindrical groove over the cuboidal one with linear polarisation becomes significant once manufacturing methods are factored in. Laser Drilling, especially with flat-top profiles inherently leads to cylindrical geometries and thus, the increased cut-off becomes a bonus advantage. With these factors in mind, it is certain that application relevant proton energies may be achieved via grooved hydrocarbon targets.

\section*{Acknowledgments}
    The authors acknowledge the EPOCH consortium for providing access to the EPOCH framework.\\
    We would also like to acknowledge the high performance computing (HPC) facility provided by Indian Institute of Technology Delhi.\\
    M.Y. acknowledges financial grant provided by the Indian Institute of Technology Delhi.\\
    V.S. acknowledges support from IIT Delhi Seed Grant No. IITD/Plg/Budget/2019-2020/197285 (revised as IITD/Plg/Budget/2020-2021/240786).\\
    The Scientific Colour Map Library \cite{scm_software} has been used in this study to prevent visual distortion of the data and exclusion of readers with colour-vision deficiencies \cite{Crameri2020}.

\bibliographystyle{unsrt}  

\end{document}